\documentclass[10pt,letterpaper]{article}
\usepackage[top=0.85in,left=2.75in,footskip=0.75in]{geometry}

\usepackage{amsmath,amssymb}

\usepackage{changepage}

\usepackage[utf8x]{inputenc}

\usepackage{textcomp,marvosym}

\usepackage{cite}

\usepackage{nameref,hyperref}

\usepackage[right]{lineno}

\usepackage{microtype}
\DisableLigatures[f]{encoding = *, family = * }

\usepackage[table]{xcolor}

\usepackage{array}

\newcolumntype{+}{!{\vrule width 2pt}}

\newlength\savedwidth

\raggedright
\usepackage[aboveskip=1pt,labelfont=bf,labelsep=period,justification=raggedright,singlelinecheck=off]{caption}

\makeatletter
\renewcommand{\@biblabel}[1]{\quad#1.}
\makeatother

\date{}

\usepackage{lastpage,fancyhdr,graphicx}
\usepackage{epstopdf}

\renewcommand{\linenumbers}{}
\renewcommand{\nolinenumbers}{}

\usepackage{multirow}

\usepackage[T1]{fontenc}

\usepackage{booktabs}

\begin{document}
\vspace*{0.2in}

\begin{flushleft}
{\Large
\textbf\newline{Separating Components of Attention and Surprise} 
}
\newline
\\
Per B{\ae}kgaard\textsuperscript{1*},
Michael Kai Petersen\textsuperscript{1},
Jakob Eg Larsen\textsuperscript{1},
\\
\bigskip
\textbf{1} Cognitive Systems, DTU Compute, Technical University of Denmark, Kgs. Lyngby, Denmark
\\
\bigskip

%
%





* pgba@dtu.dk

\end{flushleft}
\section*{Abstract}

Cognitive processes involved in both allocation of attention during decision making as well as surprise when making mistakes trigger release of the neurotransmitter norepinephrine, which has been shown to be correlated with an increase in pupil dilation, in turn reflecting raised levels of arousal. 
Extending earlier experiments based on the Attention Network Test (ANT), separating the neural components of alertness and spatial re-orientation from the attention involved in more demanding conflict resolution tasks, we demonstrate that these signatures of attention are so robust that they may be retrieved even when applying low cost eye tracking in an everyday mobile computing context. 
Furthermore we find that the reaction of surprise elicited when committing mistakes in a decision task, which in the neuroimaging EEG literature have been referred to as a negativity feedback error correction signal, may likewise be retrieved solely based on an increase in pupil dilation.

\linenumbers

\section*{Introduction}

The pupil provides a window into some of the processing that otherwise takes place invisibly inside the human brain. Hess and Polt~\cite{hess1960pupil},~\cite{hess1964pupil} as well as later Kahneman and Beatty~\cite{kahneman1966pupil},~\cite{beatty2000pupillary} found evidence that linked emotional and cognitive processes to pupil dilations, and Aston-Jones et al.~\cite{aston-jones1999role},~\cite{aston-jones2005integrative} and Joshi~\cite{joshi2016relationships} have provided a framework for understanding some of the anatomical processes that take place in regulating the gain of the networks involved, and why pupillary reactions are visible:

At the core, the Locus Coeruleus-Norepinephine (LC-NE) system operates in two different modes, \textit{tonic mode} that regulates the overall level of preparedness or arousal and \textit{phasic mode} that is involved in responding to task-relevant stimuli. As task difficulty increases, so will tonic mode activity, modulating the gain, which in turn leads to a increased performance and a stronger phasic response to task-relevant stimuli. If, however, the arousal system and tonic activity mode increase beyond a certain peak point, the phasic responses decrease, leading to an explanation of the classical trade-off between arousal and optimal performance first analysed by Yerkes and Dodson~\cite{yerkes1908relation}.

\begin{figure}[!h]
\centering
\includegraphics[width=1\columnwidth]{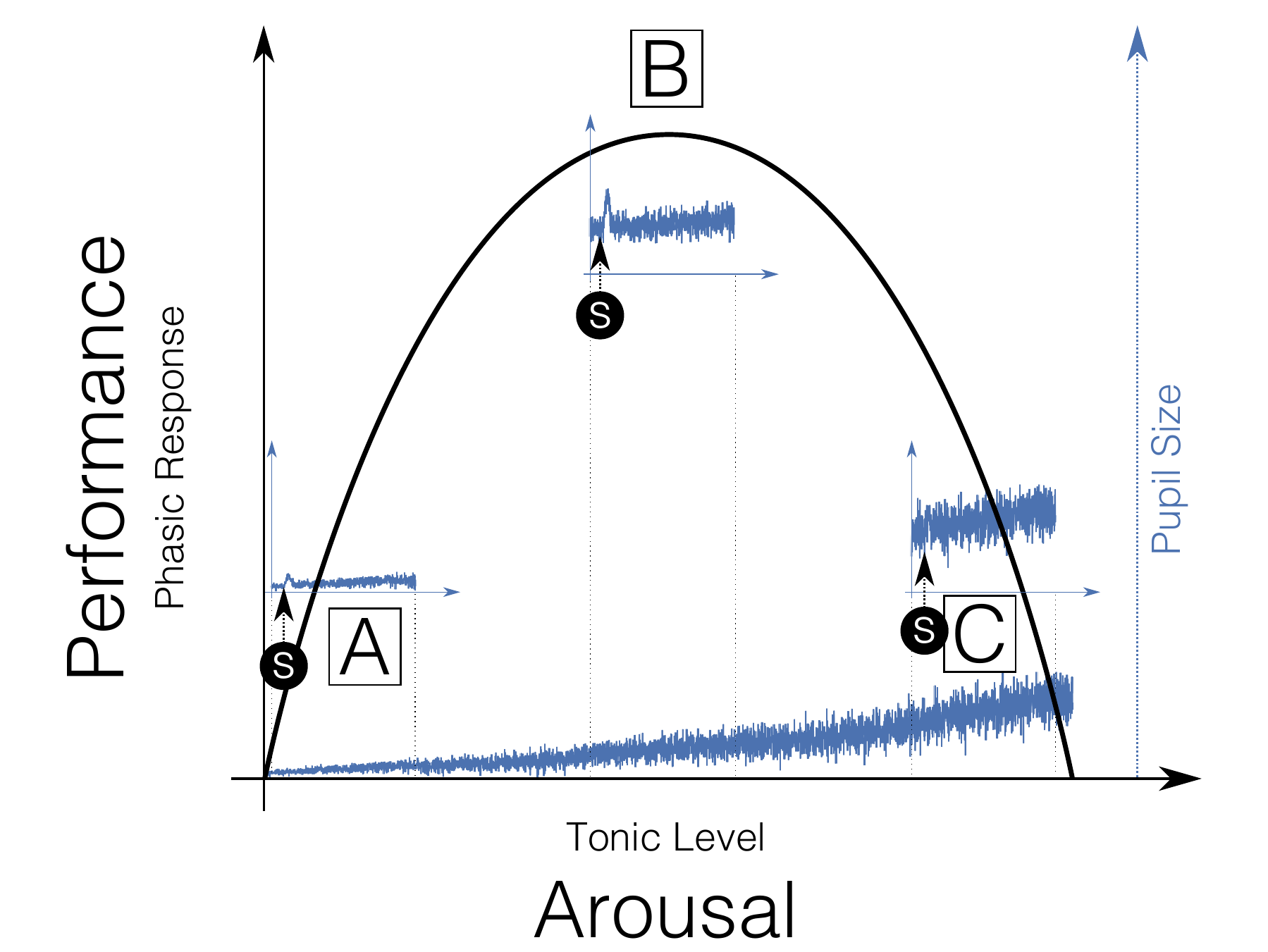}
\rule{0pt}{4ex} 
\caption{
{\bf Performance vs Tonic Level}, illustrating pupil dilations resulting from a phasic response to relevant stimuli vs the tonic baseline level, regulated by the LC-NE system. Sensitivity to task specific relevant stimuli is greatest at [B], where the largest phasic dilations are seen as compared to at [A] and [C]. Note that the graphs are not actual data to scale but is drawn for illustrative purposes.
(Adapted from~\cite{aston-jones1999role}~\cite{joshi2016relationships}, resembling the classical Yerkes-Dodson relationship~\cite{yerkes1908relation}
)}
\label{fig:phasictonic}
\end{figure}

\begin{figure}[!h]
\centering
\includegraphics[width=1\columnwidth]{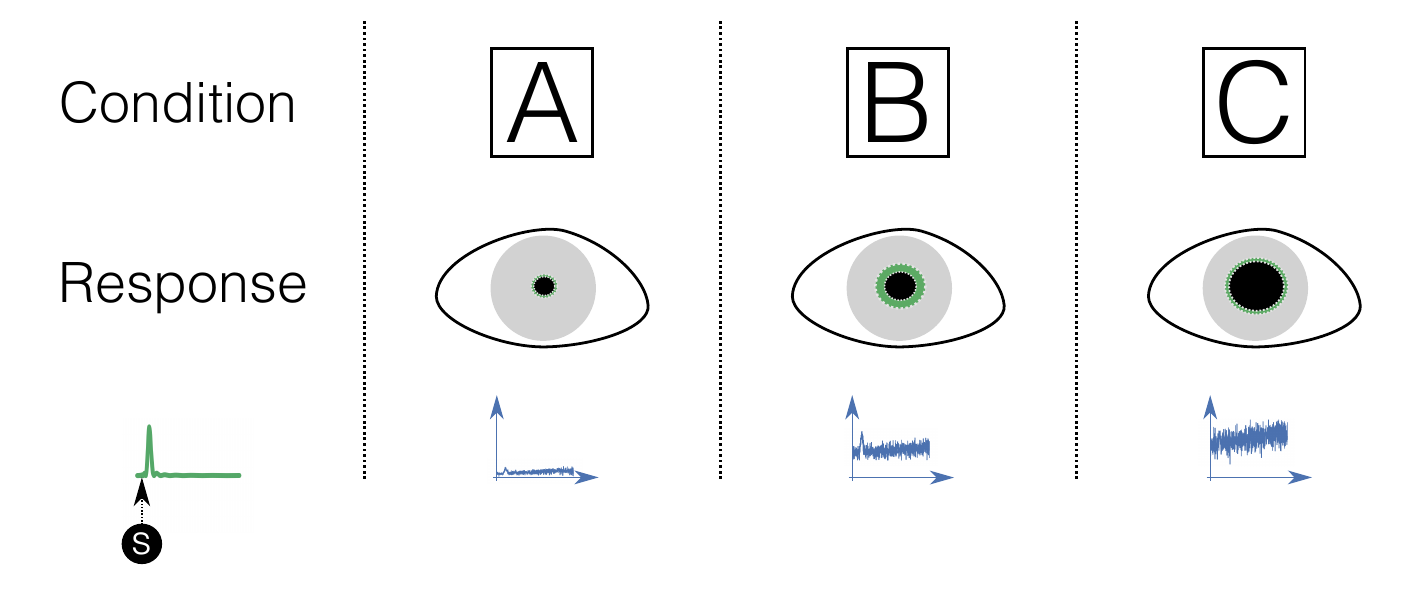}
\rule{0pt}{4ex} 
\caption{
{\bf Pupil dilation baseline vs respose} to relevant stimuli in 3 different conditions, corresponding to drowsiness [A], highly focused task-specific attention [B] and distractible, scanning attention [C]. The blue curve illustrate the level and fluctuations of the pupil size at each condition. The baseline pupil size is shown in black, with the green area denoting the size of the response present when a task-relevant stimuli appear. Note that the drawing is not to scale.
}
\label{fig:eyedilation}
\end{figure}

Activity in LC-NE cells are further reflected in pupillary dilations~\cite{joshi2016relationships}, and the pupil can thus be
interpreted as a marker of LC-NE activity (see also Fig.~\ref{fig:phasictonic} and Fig.~\ref{fig:eyedilation}). 
Baseline pupil size varies on a large scale of 3-4mm as a response to changes in light levels~\cite{walker1990clinical},~\cite{ellis1981pupillary} whereas variations caused by cognitive processes are much smaller, typically on the order of 0.5mm or 15\% compared to typical pupil sizes found in normal conditions~\cite{beatty2000pupillary}.

The baseline pupil size is modulated by the tonic activity in LC-NE, and is never at rest; it has been known for a long time to vary. Stark et al.~\cite{stark1958pupil} speculated that this could be part of an ``economical'' construction of the eye in the sense that there is no need for the eye to  operate at a more narrow range, and in our previous study~\cite{baekgaard2016assessing} we also noted slow variations of the baseline pupil size of +/-10\% on a timescale of 30--60s. 
Task-Evoked Pupillary Responses~\cite{beatty1982task},~\cite{ahern1979pupillary} (TEPR) above the current baseline are caused by phasic activity in the LC-NE system, and by averaging over many similarly conditioned tests time-locked to the stimuli, other factors can be filtered out.


Recent fMRI studies by Kuchinsky et al. have further established that activity in saliency networks triggered by attentional tasks are reflected in increased tonic pupil size, in contrast to the decreased pupil dilation typically observed when we are in a default resting state \cite{kuchinsky2016linking}.

Phasic activations of the LC-NE system in the noradrenergic (NE) neurons also play a role in rapid adaptation to changing conditions, as demonstrated by Bouret et al.~\cite{bouret2005network}, in that it may facilitate reorganisation of the innervated areas. This allows for adaption of behaviour to changes in task conditions; real or when they deviate from anticipation.


Preuschoff et al.~\cite{preuschoff2011pupil} have found that pupil dilations not only reflect decision making per se or the level of engagement, but also indicates surprise when committing mistakes in decision tasks, suggesting that NE plays a role in error signalling. This appear similar to the negativity feedback components, which is an Event Related Potential (ERP) typically observed in EEG neuroimaging experiments 250-300 ms after participants realize that an incorrect choice was made~\cite{sato2005effects}.



\section*{Materials and Methods}

While attention can be broadly understood as ``the appropriate allocation of processing resources to relevant stimuli''~\cite{coull1998neural}, Posner and Petersen~\cite{posner1990attention},~\cite{posner2012attentional},~\cite{petersen2012attention} have shown that three systems, which regulate attention, are anatomically separate from other processing systems and carry out different cognitive roles as part of the attention networks. These are:
\begin{itemize}
\item Alerting,
\item Orienting and
\item Executive Control.
\end{itemize}
Fan et al.~\cite{fan2002testing} designed a behavioral experiment, known as the Attention Network Test (ANT), to assess which of the network components are activated based on differences in reaction time when responding to visual cues.

We have in a previous experiment~\cite{baekgaard2016assessing}, measured Task-Evoked Pupillar Responses during the ANT test in a longitudinal study of two subjects. A stronger response was triggered by incongruent conditions in the conflict resolution decision task, likely involving the executive network.

This study expands the number of subjects, investigates the changes in mean pupil size over the experiment, and look at the relationship between the tonic level and the accuracy of the responses.

\subsection*{Experimental Procedure}

\begin{figure}[h!]
\def\svgwidth{\textwidth}
{\sffamily
\small
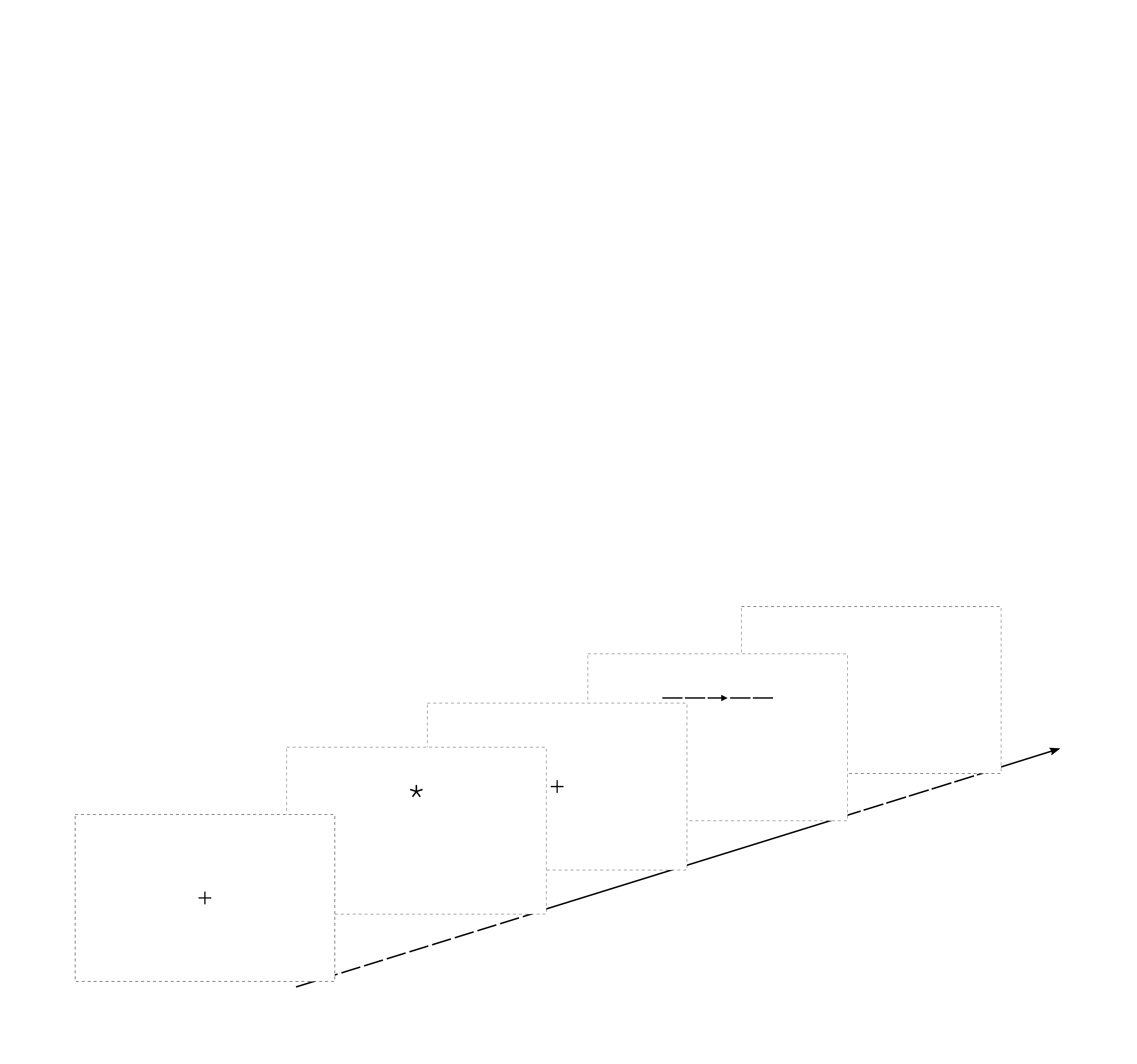
}\caption{
The Attention Network Test procedure used here: Every 4 seconds, a cue (either of 4 conditions (\textsc{Top, Left})) precedes a target (either of 3 congruency conditions (\textsc{Top, Right})), to which the participant responds by pressing a key according to the central arrow. 
The reaction time differences between cue- and congruency conditions form the basis for calculating the latencies of the attention, orientation and conflict resolution networks.
\textit{This figure and description is from \cite{baekgaard2016assessing}.}
}
\label{fig:ant-run-exp}
\end{figure}
\noindent The procedure followed and the equipment used is identical to that described in~\cite{baekgaard2016assessing}, and is further illustrated in Fig.~\ref{fig:ant-run-exp}. In this present study, in total N=18 participants (7 female and 11 male) with a mean age of 25.3 years were tested once. None used glasses or contact lenses, and all but one were right-handed.

The participant were all volunteers that were only allowed to complete the test if they gave consent to their data being 
used anonymously for research purposes. The ANT test itself is a standard paradigm in widespread use.

\subsection*{Analysis}

Pupil size is recorded at 60 Hz, and blink-affected periods are removed. A Hampel~\cite{hampel1974influence} filter with a centered window of +/-83ms and a limit of $3\sigma$ is applied to remove outliers, and when data is not present in at least half the window, the center point is also removed. This later part takes care of removing any samples immediately before and after blinks, to avoid accidental pupil size changes caused by distortion of the visible part of the eye. Finally data is downsampled to 100ms resolution with a windowed averaging filter.

For the TEPR calcuation, data is epoch'ed to the cue presentation and individually scaled to the value at the start of the epoch.

For the tonic pupil size, a period of 1s immediately before target presentation is sampled to give a representative value without the phasic response, that in most conditions appear to fade away after around 2.5s after stimuli. The pupil was further corrected for variations in head-distance by means of the eye-to-eye distance reported by the eye-tracker.

The mean pupil size was calculated in each of the 4 periods (the initial trial round and the three actual blocks of reaction time tests) as the average value of the filtered pupil data corrected for head-distance variations, which means it is representative of both the tonic pupil size and the ovelaid phasic responses.

\section*{Results}

\begin{table}[h]
\centering
\caption{Average Reaction- and Attention Network-Times over all correct tests across all users ($\pm$~Sample Standard Deviation listed in parenthesis), in seconds.}
\rule{0pt}{6ex}  
\label{tab:rtant}
\begin{tabular}{@{}cccc@{}}
\toprule
Mean\textsc{rt} & Alerting & Orienting & Conflict \\ \midrule
$0.505$ ($\pm0.074$) & $0.033$ ($\pm0.022$) & $0.019$ ($\pm0.016$) & $0.093$ ($\pm0.033$) \\ \bottomrule
\end{tabular}
\end{table}

\noindent The mean reaction time and the effect of the alerting, orienting and conflict resolution networks are summarized in table~\ref{tab:rtant}. The mean alerting effect was 33ms ($\pm$22ms), comparable to the 47ms ($\pm$18ms) reported by Fan et al.~\cite{fan2002testing}. The mean orienting and conflict effecs were 19ms ($\pm$16ms) and 93ms ($\pm$33ms) respectively, to be compared against somewhat higher 51ms ($\pm$21ms) and more similar 84 ms ($\pm$25ms).

\subsection*{Phasic Pupil Dilation vs Condition and Surprise}

\begin{figure}[!h]
\centering
\includegraphics[width=1\columnwidth]{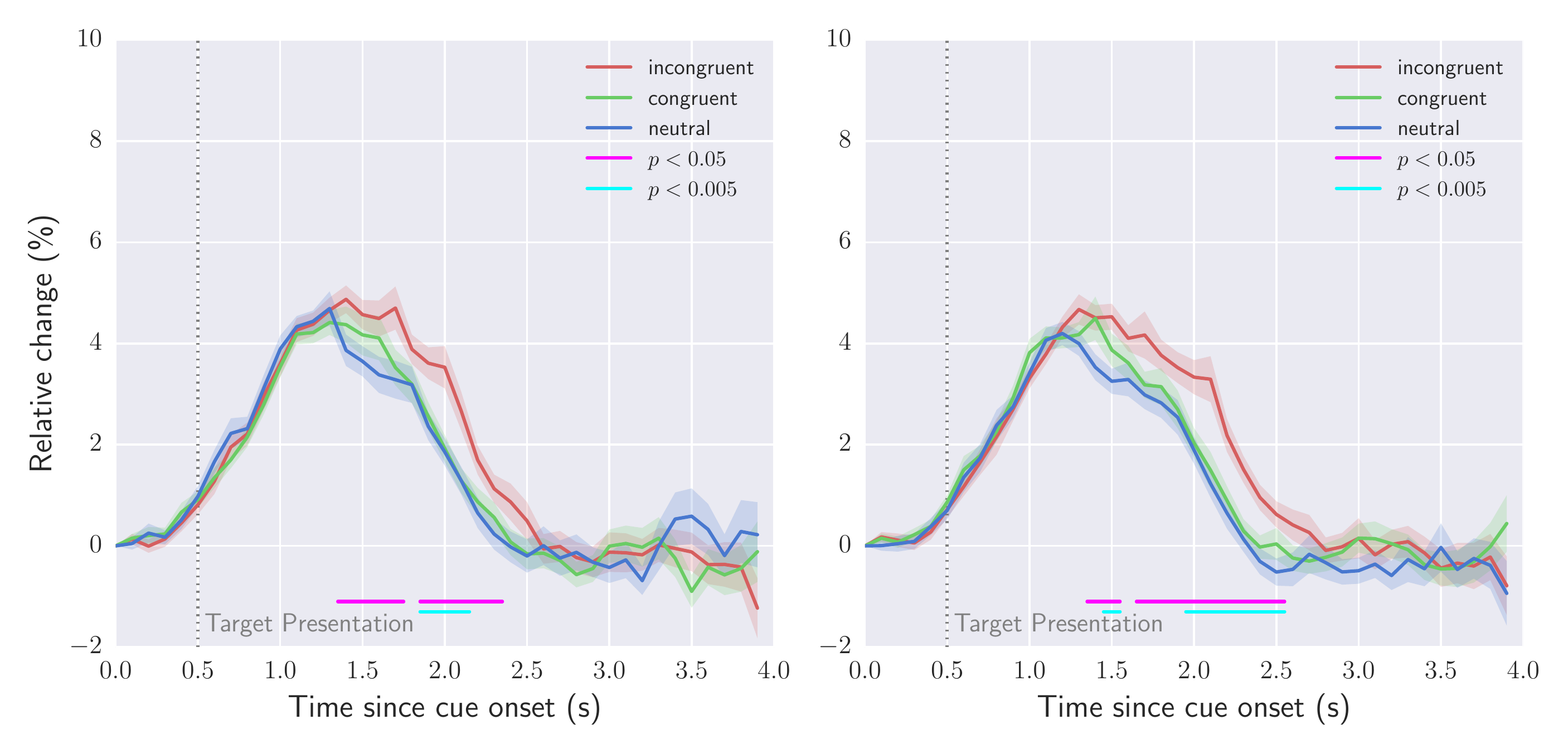}
\caption{
{\bf Average Pupil Dilation at the three congruencies} for left and right eye, respectively, for the N=18 subjects of the present study (correct responses only).
A slight initial reaction appear to the presentation of the cue at $t=0$, followed by a continued and stronger response to presentation of the target at $t=0.5$.
The congruent (green) and neutral (blue) pupil dilations are similar, but the incongruent (red) response is stronger and lasts longer.
The shaded areas represent one standard error of the mean (SEM) to each side. 
As blinks are more frequently occuring somewhere in the range of $t$ between $1$ and $2$ seconds, the SEM is somewhat larger here.
The magenta line shows where a Welch t-test between the incongruent and the neutral conditions are significantly different with a confidence level of $p<0.05$; the cyan line marks the $p<0.005$ level.
}
\label{fig:pupil_all_subjects_vs_congruency}
\end{figure}

Fig.~\ref{fig:pupil_all_subjects_vs_congruency} shows the average pupil dilation for all correctly replied tests at the three congruencies as an average value of all correctly replied tests for all subjects, for both left and right eye. The incongruent condition, where the executive control network is also invoked, shows a longer-lasting response in both eyes. Left and right eye responses do not appear statistically significant.

\begin{figure}[!h]
\centering
\includegraphics[width=1\columnwidth]{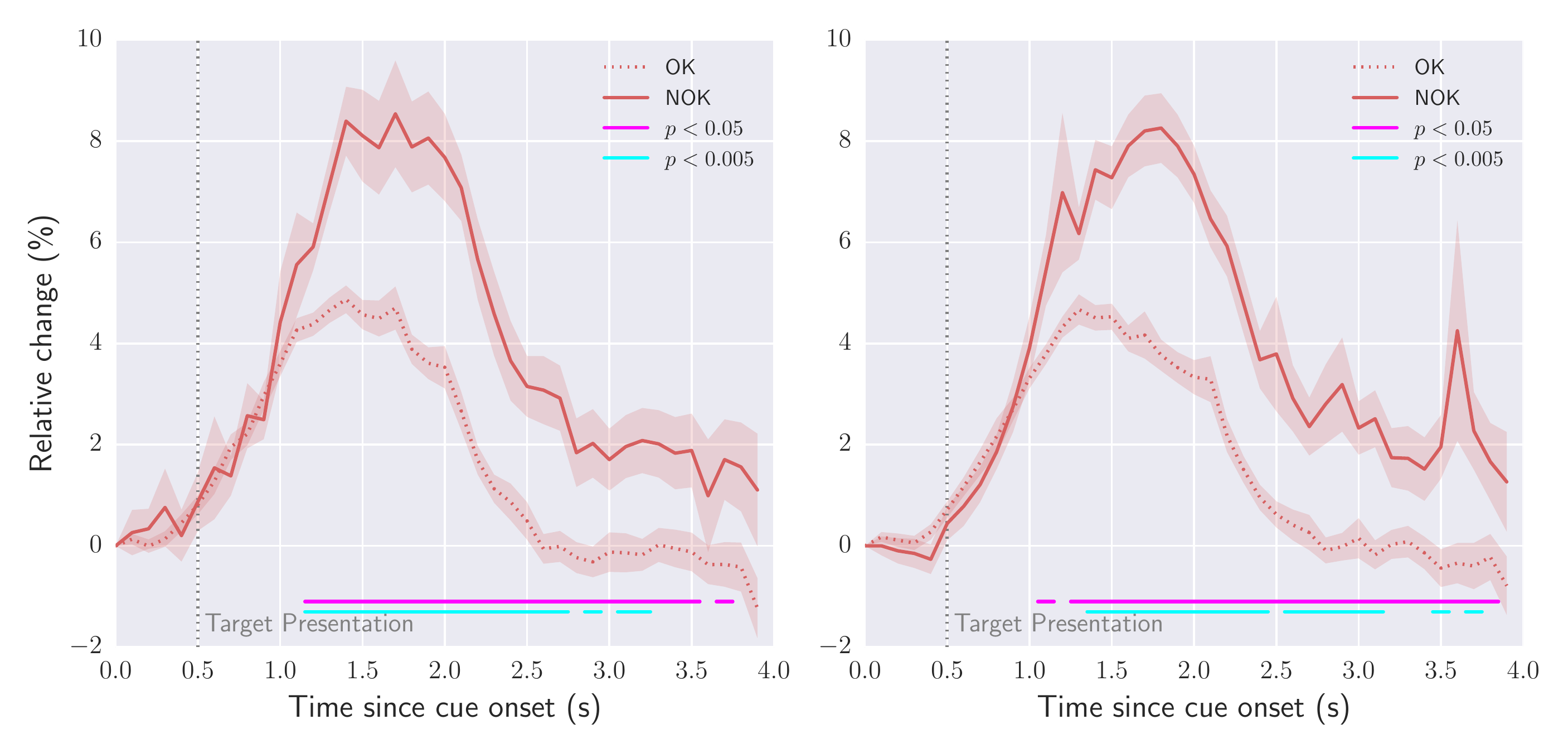}
\caption{
{\bf Average incorrect incongruent Pupil Dilations} vs correct responses for left and right eye, respectively, for the N=18 subjects, graphed similarly to Fig~\ref{fig:pupil_all_subjects_vs_congruency}.
An incorrect response invokes a statistically significant stronger pupil dilation, almost twice as large as for correct responses.
}
\label{fig:pupil_all_subjects_incongruent_vs_OK_NOK}
\end{figure}

Fig.~\ref{fig:pupil_all_subjects_incongruent_vs_OK_NOK} shows the difference between correct and incorrect (incongruent) responses. A statistically significant stronger response is seen when an error is made, which indicate that the subjects are aware of having made an error. Similar results are seen also for the other congruency conditions, and is also seen for the original longitudinal study when analysed in the same way (not shown).

\subsection*{Tonic Pupil Dilation vs Accuracy and Reaction Time}

\begin{figure}[!h]
\centering
\includegraphics[width=1\columnwidth]{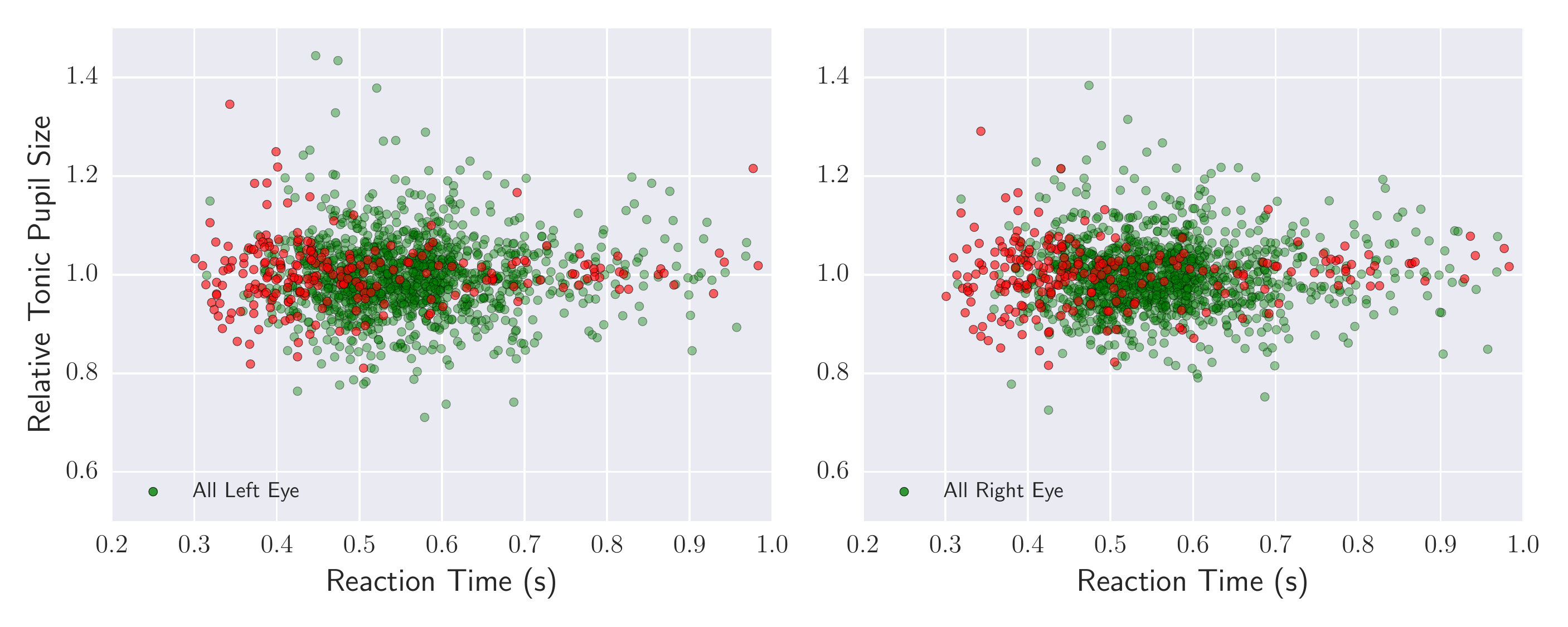}
\caption{
{\bf Scatter plot of the tonic pupil size relative to the session mean vs reaction time} for incongruent conditions for all participants (left and right eye), colour coded according to correct (green) and incorrect (red) responses.
The mean reaction time between correct and incorrect responses are significantly different for the incongruent condition (Welch t-test $t=7.00$, $p<0.000001$).
The mean relative tonic pupil size between correct and incorrect responses do not significantly.
}
\label{fig:pd_vs_rt_all_incongruent_both}
\end{figure}

\begin{figure}[!h]
\centering
\includegraphics[width=1\columnwidth]{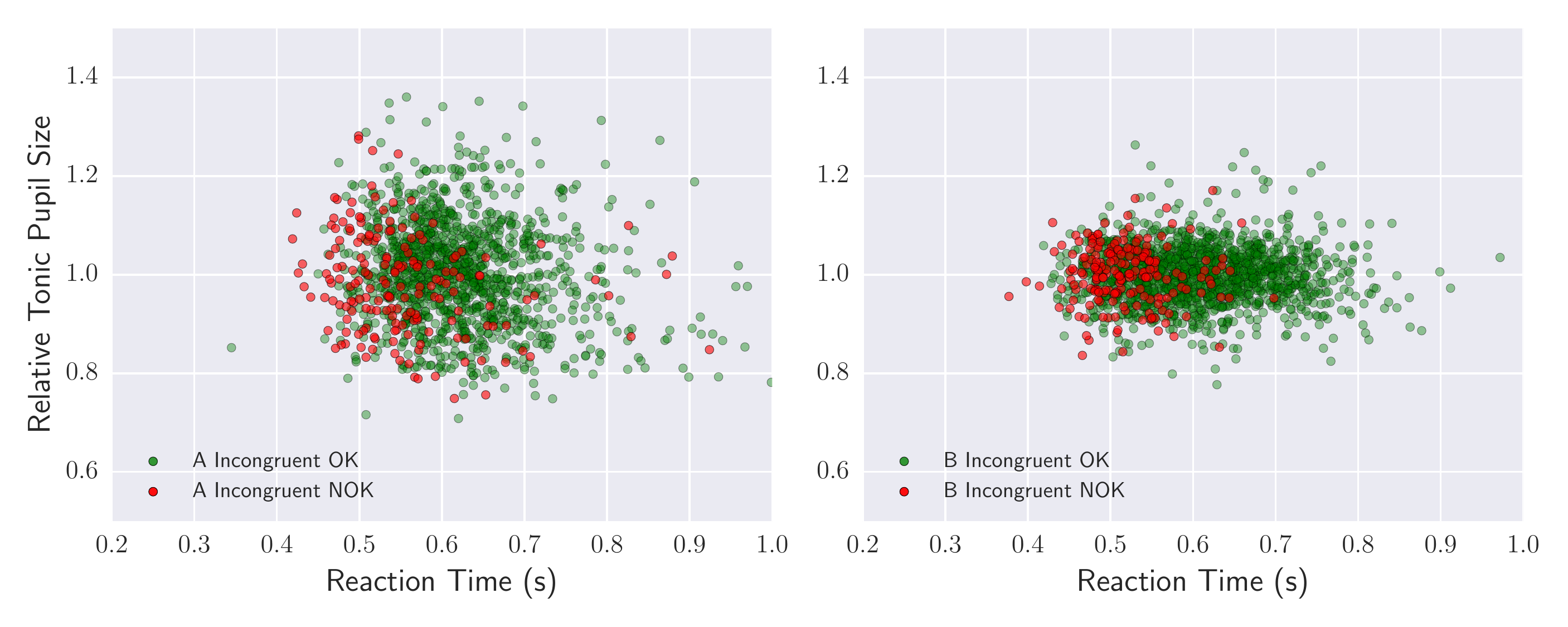}
\caption{
{\bf Scatter plot of the left eye tonic pupil size relative to the session mean vs reaction time} for incongruent conditions for both participant A (left) and B (right) over all sessions of the longitudinal study, colour coded according to correct (green) and incorrect (red) responses (left eye only shown).
The mean reaction time between correct and incorrect responses are significantly different for the incongruent condition for either participant; see Table~\ref{tab:psz_rt}.
The mean relative tonic pupil size between correct and incorrect responses only differ significantly for participant A (Welch t-test $t=2.47$, $p=0.014$ left eye and $t=2.39$, $p=0.017$ right eye); for B they appear very similar.
Also note that variations in the relative tonic pupil size appear larger for A than for B.
}
\label{fig:pd_vs_rt_a_b_incongruent}
\end{figure}

Fig.~\ref{fig:pd_vs_rt_all_incongruent_both} and Fig.~\ref{fig:pd_vs_rt_a_b_incongruent} show scatter plots of all correct (green) and incorrect (red) responses to the incongruent condition according to the reaction time and the tonic pupil size immediately before the test. Incorrect replies are associated with shorter reaction times. 
The reaction times are statistically different between conditions for all participants of the present study and for A and B in the longitudinal study.
The tonic pupil size does not differ in the present study between conditions. However, for the original longitudinal study, participant A shows a statistically significant difference between conditions, with the mean tonic pupil size smaller when incorrect replies are given. See also Table~\ref{tab:psz_rt}.

\begin{table}[!ht]
\caption{
{\bf Relative tonic pupil size and reaction time}
}
\centering
\label{tab:psz_rt}
\begin{tabular}{lllrrrr}
                  &                           &                              & \multicolumn{1}{c}{$\mu$} & \multicolumn{1}{c}{SEM} & \multicolumn{1}{c}{N} & \multicolumn{1}{c}{p()} \\ \hline
\multirow{6}{*}{All (N=18)} & \multirow{3}{*}{PSz}       & OK                           & 0.998                     & 0.002                   & 1472                  &                         \\
                  &                           & NOK                          & 1.010                     & 0.007                   & 254                   &                         \\
                  &                           & \multicolumn{1}{r}{$\delta$} & 0.017                     &                         &                       & 0.116                   \\ \cline{2-7} 
                  & \multirow{3}{*}{$\mu RT$} & OK                           & 0.568                     & 0.003                   & 1472                  &                         \\
                  &                           & NOK                          & 0.496                     & 0.010                   & 254                   &                         \\
                  &                           & \multicolumn{1}{r}{$\delta$} & -0.072                    &                         &                       & 0.000                   \\ \hline
\multirow{6}{*}{A} & \multirow{3}{*}{PSz}       & OK                           & 1.003                     & 0.003                   & 1333                  &                         \\
                  &                           & NOK                          & 0.982                     & 0.007                   & 198                   &                         \\
                  &                           & \multicolumn{1}{r}{$\delta$} & -0.021                    &                         &                       & 0.014                   \\ \cline{2-7} 
                  & \multirow{3}{*}{$\mu RT$} & OK                           & 0.631                     & 0.003                   & 1333                  &                         \\
                  &                           & NOK                          & 0.572                     & 0.009                   & 198                   &                         \\
                  &                           & \multicolumn{1}{r}{$\delta$} & -0.059                    &                         &                       & 0.000                   \\ \hline
\multirow{6}{*}{B} & \multirow{3}{*}{PSz}       & OK                           & 1.000                     & 0.002                   & 1434                  &                         \\
                  &                           & NOK                          & 1.001                     & 0.004                   & 197                   &                         \\
                  &                           & \multicolumn{1}{r}{$\delta$} & 0.001                     &                         &                       & 0.845                   \\ \cline{2-7} 
                  & \multirow{3}{*}{$\mu RT$} & OK                           & 0.612                     & 0.002                   & 1434                  &                         \\
                  &                           & NOK                          & 0.519                     & 0.004                   & 197                   &                         \\
                  &                           & \multicolumn{1}{r}{$\delta$} & -0.093                    &                         &                       & 0.000                   \\ \hline

\end{tabular}

\begin{flushleft} Left eye tonic pupil size, as measured immediately before target presentation, relative to each session's mean, and the reaction times, are listed across all subjects of the present study, and for both participant A and B over all sessions of the longitudinal study, for incongruent conditions, divided into groups of correct and incorrect responses.
The mean reaction time ($\mu RT$) differ between correct and incorrect responses in a significant way (Welch' t-test $t=7.00$, $t=5.99$ and $t=21.29$ respectively, p<0.000001) for both A and B.
The means of the tonic pupil size (PSz) differ significantly between correct and incorrect responses for A (Welch' t-test $t=2.47$, $p=0.014$); for B and the participants of the present study, the means between the conditions do not show a statistically significant difference.
Almost identical results are found for right eye pupil sizes (not listed here).
\end{flushleft}
\label{tab:ok_vs_nok_incongruent}
\end{table}

\subsection*{Tonic Pupil Dilation over time}

\begin{figure}[!h]
\centering
\includegraphics[width=1\columnwidth]{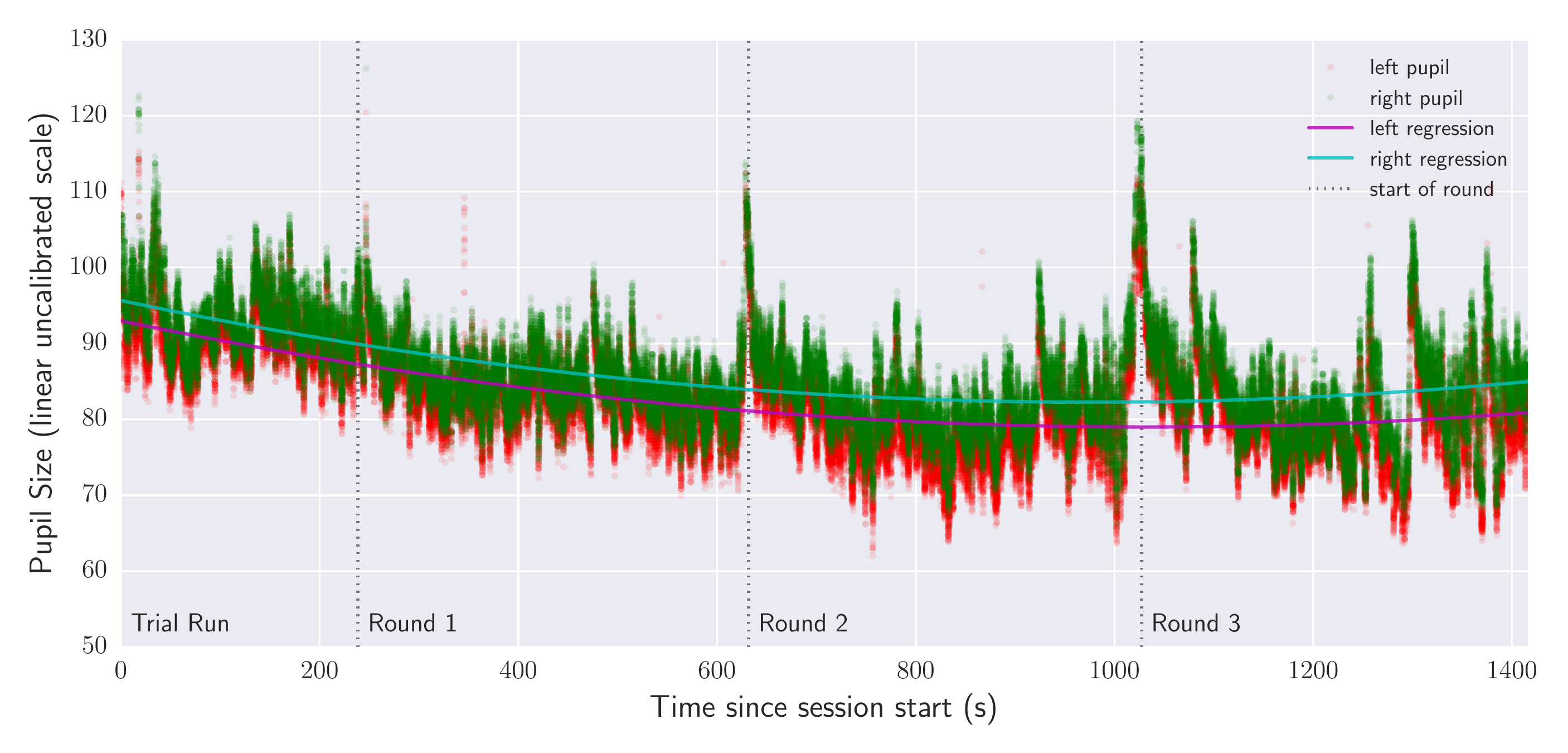}
\caption{
{\bf Pupil Size over a sample session that illustrates interesting trends clearly.} 
This session is a first-run for a participant in the larger study. Red and Green marks the measured pupil size (compensated for changes in head distance) for left and right eye respectively. 
The solid magenta and cyan lines are 2nd order approximations to the pupil size; the explained variance ($R^2$) are 0.345 and 0.301 respectively.
Initiation of each of the 3 rounds of the session are marked with dashed lines.
An initial increased pupil dilation diminishes over time as entraining takes place, with a slight increase towards the end.
It can also be seen that, in this case, each round starts out with an increase pupil dilation.
}
\label{fig:tonic_pupil_20160112150851}
\end{figure}

Fig.~\ref{fig:tonic_pupil_20160112150851} shows an illustrative sample of how the mean pupil size (corrected for variations in head distance) varies over the course of the initial training round and the three actual trial blocks. Left and right pupil size are slightly different for this particular subject, but there is good correlation between variations of the two (Pearson's $R=0.948$). An regression corresponding to a low pass filter (a 2nd order polynomial fit) is shown overlaid, and can explain approximately 30-35\% of variance (explained variance $R^2=0.345$ and $R^2=0.301$ respectively).

It also appears as if each block has a slightly larger tonic pupil size initially followed by a decline of approximately 10\%. The means for each block also apper to differ, with the initial training round having the larger tonic pupil size.

\begin{figure}[!h]
\centering
\includegraphics[width=1\columnwidth]{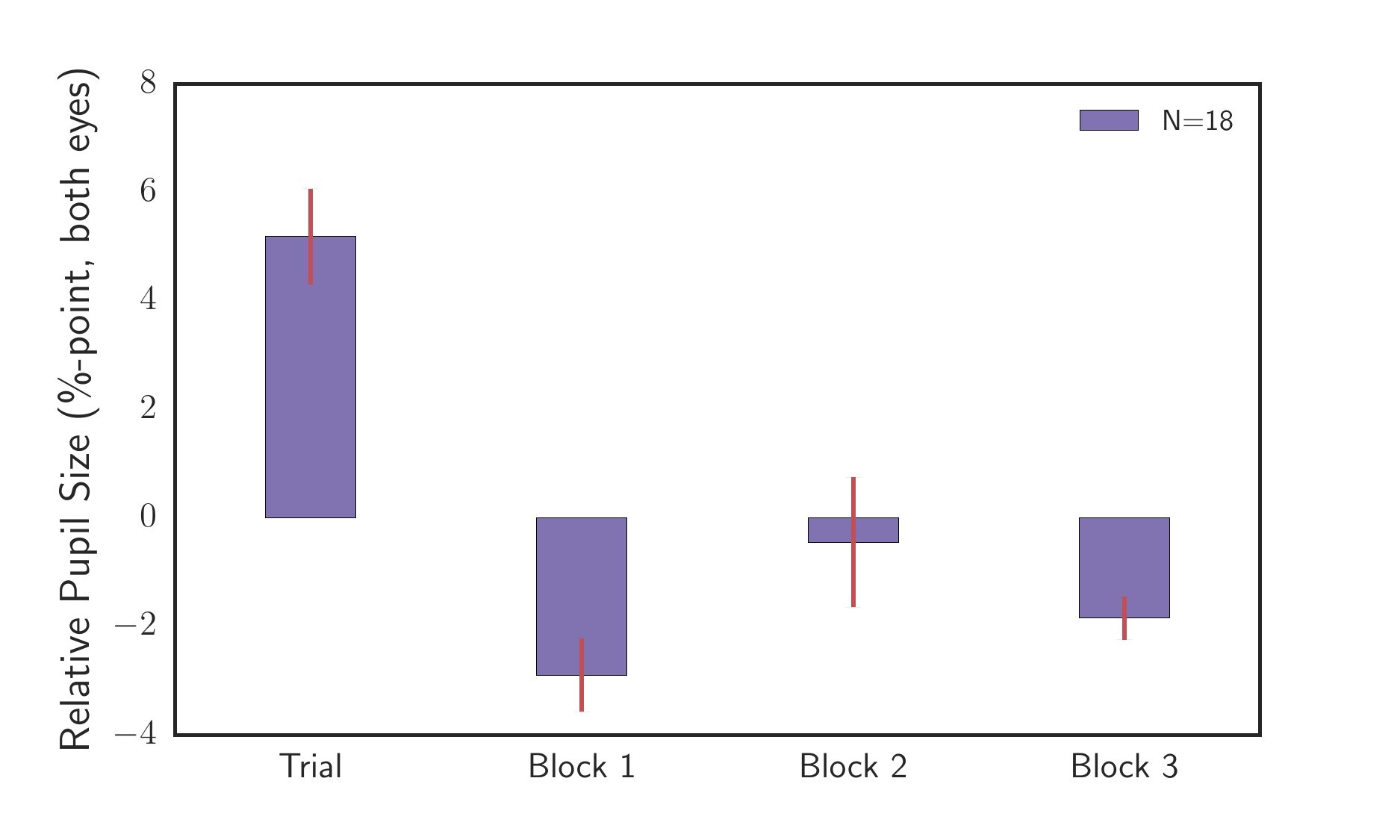}
\caption{
{\bf Mean Relative Pupil Size for all subjects in this present study} divided into the initial training round and the three actual trial blocks.
The red line denotes the SEM.
The differences between the initial training round and any of the three other blocks are statistically significant (Wilcoxon signed-rank test $T=12$, $T=111$, $T=35$, all with a confidence level $p<0.001$). The differences between the other blocks are not statistically significant.
}
\label{fig:qmeans_all}
\end{figure}

\begin{figure}[!h]
\centering
\includegraphics[width=1\columnwidth]{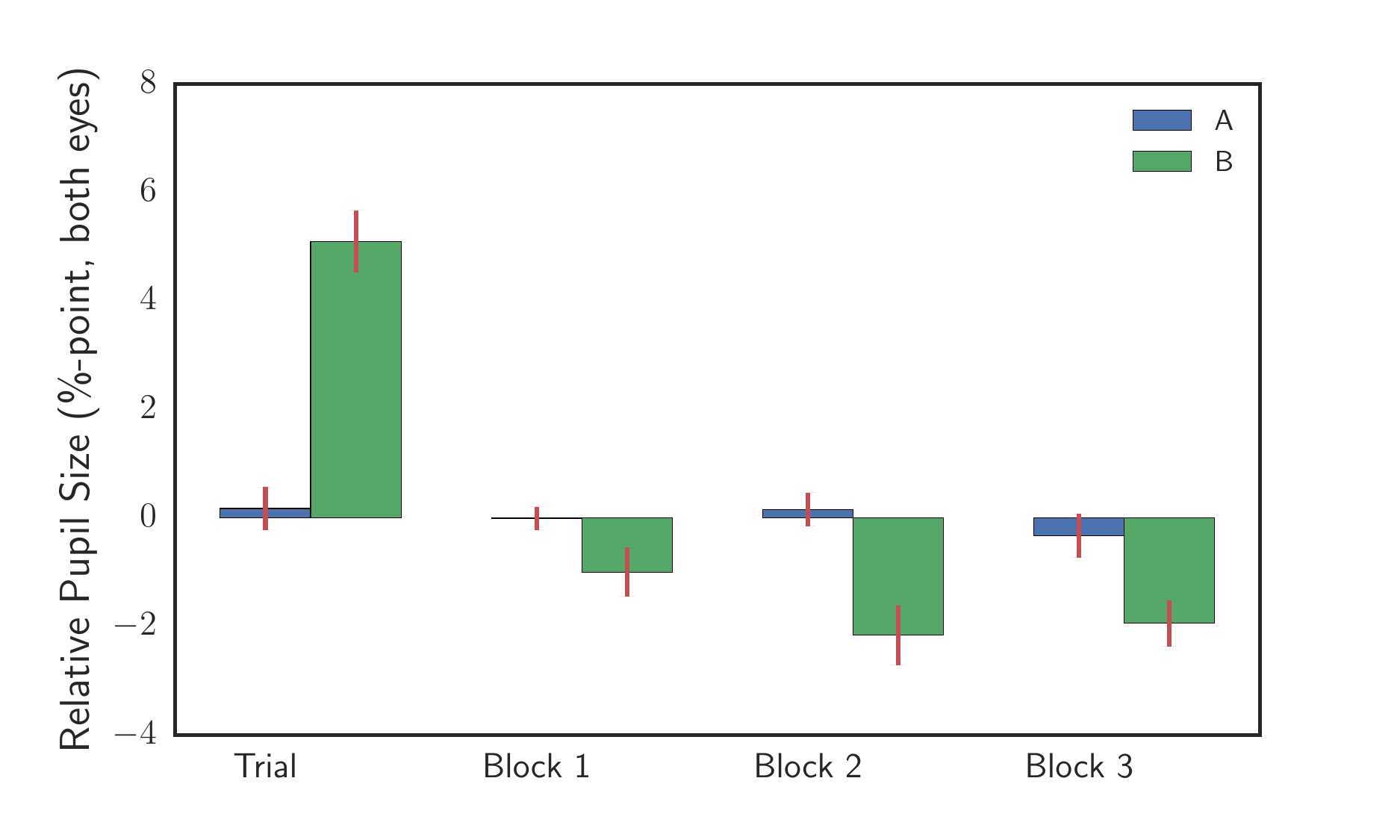}
\caption{
{\bf Mean Relative Pupil Size for all subjects in the longitudinal study} divided into the initial training round and the three actual trial blocks.
The red line denotes the SEM.
The differences between the initial training round and any of the three other blocks are statistically significant for B (Wilcoxon signed-rank test $T=18$, $T=13$, $T=18$, all with a confidence level $p<0.001$). The differences between the other blocks are not statistically significant.
However, for A there are no statistically significant differences between the blocks; the variations between the 4 blocks are comparatively much smaller than than what is seen for other participants.
}
\label{fig:qmeans_ab}
\end{figure}

When comparing the mean tonic pupil size between the initial training round and the three actual trial blocks, there are statistically significant differences across all participants of the present study, and also for participant B of the longitudinal study. For participant A, however, there are no statistically significant differences. See Fig.~\ref{fig:qmeans_all} and  Fig.~\ref{fig:qmeans_ab}

\subsection*{Fixation Density Map Differences}
  
Average Fixation Density Maps, adjusted for accidental mis-calibrations, were built for each experiment, and were compared between conditions. 
We did, as expected, see recognisable differences when the target presentation was above vs below the fixation cross, but we were not able to detect any significant spatial differences between congruency conditions nor between cue conditions.

\section*{Discussion}


The results of this study, with a larger population, supports our previous findings: There is a difference in 
the incongruent vs congruent/neutral flanker scenarios in that an incongruent condition solicit a larger
pupillary response compare to the other two conditions.
As the age group is different compared to the previous study, there are indications that the results may be
robust and can translate to different settings..

In most cases we see a high correlation (R values from 0.8--0.95) between left and right 
pupil size, although a few have what may be less than optimal tracking. We cannot conclude any significant 
difference in the pupil dilation responses between the two eyes, but we notice that the significance level of the 
difference between the incongruent and the neutral condition is higher and lasts slightly longer for the right
eye.

In addition, we also found a significantly different response when subjects replied incorrectly, which happens much more
frequently for the inconguent condition.
This response may be related to the adaptation and required reorganization reported by Bouret et al.~\cite{bouret2005network} 
and/or to the surprise elements reported by Preuschoff et al.~~\cite{preuschoff2011pupil}.

Thus, the phasic response reported here as well as in our previous study can be divided into two components 
that apparently cause a higher level of LC-NE activations: one related to the incongruent condition and one to 
the incorrect reply.

Comparing the mean relative pupil size over the 4 parts of the experiment (training round and 3 blocks of tests)
we found that for the subjects of the present study, as well as for subject B of the longitudinal study, the
training round had a statistically significant higher level, around 5\%, compared to the other three blocks that averaged
around -2\%. The subject A of the longitudinal study, however, did not show any such variation between the blocks.
We hypothesize that this may point to differences in individual characteristics, behavoiur or preferences.

Further, comparing relative normalized tonic pupil sizes (excluding the phasic responses) showed a statistically significant
difference between the level immediately before an incorrect reply compared to the level before a correct reply
for subject A of the longitudinal study but not for subject B nor for the participants of the present study.
However, while for subject B the levels are almost identical, there is a larger difference even if not statistically
significant for the participants of this study, and it therefore cannot be ruled out that participants could fall in 
different groups that, with more data, would reveal more individual variation.

We also point out that the possible familiarity effects of higher pupillary responses mainly in the two 
first complete experiments were not tested for in the present experiment, since it was performed only once for each participant.
We do, however, see hints at an overall adaptation, as the average (tonic) level decreases as initial entraining to the tasks 
take place, with a flat or in some cases slightly increased tonic levels towards task completion.
This appear similar to the \textit{familiarity effect} reported by Hy\"on\"a et al.~\cite{hyona1995pupil}

We were not able to find any spatial differences in eye movements, at the resolution we worked with, that was related to 
the conditions of the test, apart from the up-down position of the target.



\section*{Acknowledgments}

This work is supported in part by the Innovation Fund Denmark through the project Eye Tracking for Mobile Devices.

\nolinenumbers

%
%
%

\begin{thebibliography}{10}

\bibitem{hess1960pupil}
Hess EH, Polt JM.
\newblock {Pupil size as related to interest value of visual stimuli}.
\newblock Science. 1960;132:349--50.
\newblock Available from:
  \url{http://www.ncbi.nlm.nih.gov/entrez/query.fcgi?cmd=Retrieve{\&}db=PubMed{\&}dopt=Citation{\&}list{\_}uids=14401489}.

\bibitem{hess1964pupil}
Hess EH, Polt JM.
\newblock {Pupil Size in Relation to Mental Activity during Simple
  Problem-Solving}.
\newblock Science. 1964;143(3611):1190--1192.
\newblock Available from:
  \url{http://www.jstor.org.proxy.findit.dtu.dk/stable/1712692}.

\bibitem{kahneman1966pupil}
Kahneman D, Beatty J.
\newblock {Pupil Diameter and Load on Memory}.
\newblock Science. 1966;154(3756):1583--1585.
\newblock Available from:
  \url{http://www.jstor.org.proxy.findit.dtu.dk/stable/1720478}.

\bibitem{beatty2000pupillary}
Beatty J, Lucero-Wagoner B.
\newblock {The Pupillary System}.
\newblock Cacioppo JT, Tassinary L, Berntson G, editors. Cambridge University
  Press; 2000.

\bibitem{aston-jones1999role}
Aston-Jones G, Rajkowski J, Cohen J.
\newblock {Role of locus coeruleus in attention and behavioral flexibility}.
\newblock Biological Psychiatry. 1999;46(9):1309--1320.

\bibitem{aston-jones2005integrative}
Aston-Jones G, Cohen JD.
\newblock {An Integrative Theory of Locus Coeruleus-Norepinephrine Function:
  Adaptive Gain and Optimal Performance}.
\newblock Annual Review of Neuroscience. 2005;28(1):403--450.
\newblock Available from:
  \url{http://www.annualreviews.org/doi/abs/10.1146/annurev.neuro.28.061604.135709}.

\bibitem{joshi2016relationships}
Joshi S, Li Y, Kalwani RM, Gold JI.
\newblock {Relationships between Pupil Diameter and Neuronal Activity in the
  Locus Coeruleus, Colliculi, and Cingulate Cortex}.
\newblock Neuron. 2016;89(1):221--234.
\newblock Available from: \url{http://dx.doi.org/10.1016/j.neuron.2015.11.028}.

\bibitem{yerkes1908relation}
Yerkes RM, Dodson JD.
\newblock {The relation of strength of stimulus to rapidity of
  habit-formation}.
\newblock Journal of Comparative Neurology and Psychology. 1908;18(5):459--482.

\bibitem{walker1990clinical}
Walker HK, Hall WD, Hurst JW.
\newblock {Clinical Methods}.
\newblock Butterworths; 1990.
\newblock Available from: \url{http://www.ncbi.nlm.nih.gov/pubmed/21250045}.

\bibitem{ellis1981pupillary}
Ellis CJ.
\newblock {The pupillary light reflex in normal subjects.}
\newblock The British journal of ophthalmology. 1981;65(11):754--9.
\newblock Available from:
  \url{http://www.pubmedcentral.nih.gov/articlerender.fcgi?artid=1039657{\&}tool=pmcentrez{\&}rendertype=abstract}.

\bibitem{stark1958pupil}
Stark L, Campbell FW, Atwood J.
\newblock {Pupil Unrest: An Example of Noise in a Biological Servomechanism}.
\newblock Nature. 1958 sep;182(4639):857--858.
\newblock Available from:
  \url{http://www.nature.com/doifinder/10.1038/182857a0}.

\bibitem{baekgaard2016assessing}
B{\ae}kgaard P, Petersen MK, Larsen JE.
\newblock In: Antona M, Stephanidis C, editors. {Assessing Levels of Attention
  Using Low Cost Eye Tracking}. Cham: Springer International Publishing; 2016.
  p. 409--420.
\newblock Available from:
  \url{http://dx.doi.org/10.1007/978-3-319-40250-5{\_}39}.

\bibitem{beatty1982task}
Beatty J. {Task-evoked pupillary responses, processing load, and the structure
  of processing resources}; 1982.

\bibitem{ahern1979pupillary}
Ahern S, Beatty J.
\newblock {Pupillary responses during information processing vary with
  Scholastic Aptitude Test scores}.
\newblock Science. 1979;205(4412):1289--1292.
\newblock Available from:
  \url{http://www.sciencemag.org/cgi/doi/10.1126/science.472746}.

\bibitem{kuchinsky2016linking}
Kuchinsky SE, Pand{\v{z}}a NB, Haarmann HJ.
\newblock In: Schmorrow DD, Fidopiastis MC, editors. {Linking Indices of Tonic
  Alertness: Resting-State Pupil Dilation and Cingulo-Opercular Neural
  Activity}. Cham: Springer International Publishing; 2016. p. 218--230.
\newblock Available from:
  \url{http://dx.doi.org/10.1007/978-3-319-39955-3{\_}21}.

\bibitem{bouret2005network}
Bouret S, Sara SJ.
\newblock {Network reset: A simplified overarching theory of locus coeruleus
  noradrenaline function}.
\newblock Trends in Neurosciences. 2005;28(11):574--582.

\bibitem{preuschoff2011pupil}
Preuschoff K, {'t Hart} BM, Einh??user W.
\newblock {Pupil dilation signals surprise: Evidence for noradrenaline's role
  in decision making}.
\newblock Frontiers in Neuroscience. 2011;5(SEP):1--12.

\bibitem{sato2005effects}
Sato A, Yasuda A, Ohira H, Miyawaki K, Nishikawa M, Kumano H, et~al.
\newblock {Effects of value and reward magnitude on feedback negativity and
  P300.}
\newblock Neuroreport. 2005;16(4):407--411.

\bibitem{coull1998neural}
Coull JT. {Neural correlates of attention and arousal: Insights from
  electrophysiology, functional neuroimaging and psychopharmacology}; 1998.

\bibitem{posner1990attention}
Posner MI, Petersen SE.
\newblock {The attention system of the human brain.}
\newblock Annual review of neuroscience. 1990;13:25--42.
\newblock Available from:
  \url{http://www.pubmedcentral.nih.gov/articlerender.fcgi?artid=3413263{\&}tool=pmcentrez{\&}rendertype=abstract$\backslash$nhttp://www.ncbi.nlm.nih.gov/pubmed/2183676}.

\bibitem{posner2012attentional}
Posner MI.
\newblock {Attentional networks and consciousness}.
\newblock Frontiers in Psychology. 2012;3(MAR):1--4.
\newblock Available from:
  \url{http://www.ncbi.nlm.nih.gov/pmc/articles/PMC3298960/}.

\bibitem{petersen2012attention}
Petersen SE, Posner M.
\newblock {The Attention System of the Human Brain: 20 Years After}.
\newblock Annual review of neuroscience. 2012;21(35):73--89.

\bibitem{fan2002testing}
Fan J, McCandliss BD, Sommer T, Raz A, Posner MI.
\newblock {Testing the Efficiency and Independence of Attentional Networks}.
\newblock Journal of Cognitive Neuroscience. 2002;14(3):340--347.
\newblock Available from:
  \url{http://www.mitpressjournals.org/doi/abs/10.1162/089892902317361886}.

\bibitem{hampel1974influence}
Hampel FR.
\newblock {The Influence Curve and its Role in Robust Estimation}.
\newblock Journal of the American Statistical Association.
  1974;69(346):383--393.
\newblock Available from:
  \url{http://www.tandfonline.com/doi/abs/10.1080/01621459.1974.10482962}.

\bibitem{hyona1995pupil}
Hy{\"{o}}n{\"{a}} J, Tommola J, Alaja AM.
\newblock {Pupil Dilation as a Measure of Processing Load in Simultaneous
  Interpretation and Other Language Tasks}.
\newblock The Quarterly Journal of Experimental Psychology Section A.
  1995;48(3):598--612.
\newblock Available from:
  \url{http://www.tandfonline.com/doi/abs/10.1080/14640749508401407}.

\end{thebibliography}

\end{document}